\documentclass[twoside]{ilcws10}
\usepackage[latin1]{inputenc}
\usepackage[dvips]{graphicx,epsfig,color}
\usepackage {subfigure}
\usepackage{wrapfig,rotating}
\usepackage{amssymb,amsmath,array}

\begin{document}
\title{
Test Beam Results Using an RPC Semi-Digital HCAL
} 
\author{K.Belkadhi\\
On behalf of the CALICE SDHCAL group
\thanks{Thanks to C. Combaret, M.Bedjidian, I. Laktineh, N.Lumb, R. Kieffer, M. Vander Dockt (IPNL France), V. Boudry, D. Decotigny, M. Ruan (LLR France), M-C. Fouz, J. Puerta Pelayo (CIMAP Spain), E. Cortina, S. Manai (CP3 Belgium) and K. Manai (FST Tunisia).}
\vspace{.3cm}\\
Laboratoire Leprince-Ringuet \'Ecole Polytechnique CNRS-IN2P3 \\
Palaiseau - France}


\maketitle

\begin{abstract}
We report on the development of an GRPCs (Glass Resistive Plate Chambers) Semi-Digital hadron calorimeter for the future International Linear Collider (ILC). Two types of GRPCs (small and 1m$^2$) were tested in PS beam at CERN. Detector performances are presented here in terms of efficiency, pad multiplicity, homogeneity and stability in time.
\end{abstract}

\section{Introduction}
One of the challenges of the detectors near the future International Linear Collider is the jet energy resolution \cite{Djouadi:2007ik}. The Particle Flow Algorithms (PFA)\cite{Brient:2002gh,Thomson:2009rp} require a good particles separation in jets and a high level tracking ability in the different sub-detectors.
To satisfy these conditions, a highly granular hadron calorimeter both in longitudinal and transverse directions is needed.

A semi-digital hadron calorimeter composed of 1$\times$1 cm$^{2}$ cells and up to 48 sampling layers (yielding 7$\times$10$^{7}$ channels in the entire hadron calorimeter of a large detector such as the ILD\cite{Group:2010eu}) and using Glass Resistive Plate Chambers (GRPC) as active medium, combines excellent performances and reduced cost.

\begin{figure}[h]
\begin {center}
\includegraphics[width=0.7\textwidth]{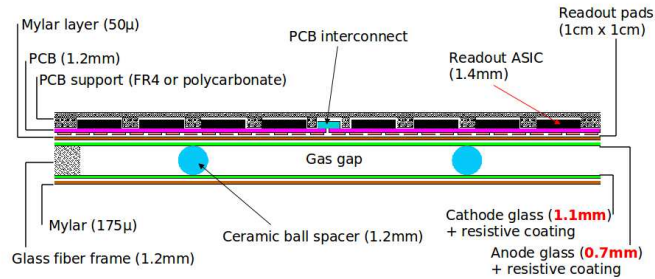}
\caption{Shematic of a standard RPC. The total thikness including embedded electronics is 5.825 mm.}
\label{Fig.RPC}
\end {center}
\end{figure}

One of the candidates for this detector is the semi-digital hadron calorimeter using GRPC's as active medium. A GRPC is a simple parallel plate sensor, using gas amplification in avalanche mode characterized by a high time and space resolution. With embedded electronics installed directly on the anode, it provides a thin sensor which satisfies the requirements of the ILC hadron calorimeter \cite{Behnke:2007gj}. A sketch of the layout is presented on Figure~\ref{Fig.RPC}.

Two types of GRPCs were built and tested in beam conditions corresponding to the different steps in the development of a cubic meter prototype. The first type is used in a set of four small GRPCs (8$\times$32 cm$^{2}$ each) used with 4 chained ASICs \cite{Bouchel:2007zz} (1 ASIC=64 channels) while the second type is a square meter GRPC read out by 144 ASICs. Their performances are presented and discussed here.

\section{The beam test setup and aquisition}

Small and 1m$^{2}$ GRPCs were exposed to PS and SPS beams at CERN during several beam tests as detailed in Table \ref{Tab.beamtests}. A mechanical structure was concieved to host the GRPCs vertically and to control the distance between them. A triggering system based on scintillators was placed in front and behind the setup.
\newpage 
The first version of ASICs (HARDROC1) used for these beam tests has a set of two thresholds and has an internal memory able to store up to 127 frames. When a trigger signal is received, the acquisition is stopped and frames stored in the ASIC memory are read. An event is a collection of frames saved with their corresponding time to the trigger and channel identifier.

\begin{figure}[!h]
\begin {center}
\includegraphics[width=0.7\textwidth]{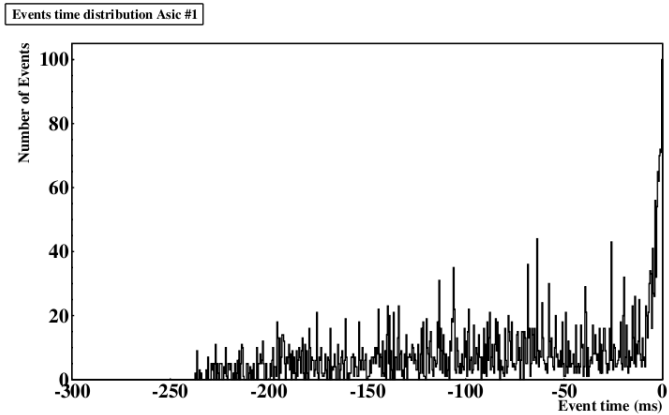}
\caption{Frames time structure for 100 triggers (1 ASIC).} 
\label{Fig.timestructure}
\end {center}
\end{figure}

 The time structure of a sample of 100 events, with the time of the frames realigned on the trigger time, realigned is shown in Figure \ref{Fig.timestructure} where the first frame in memory is the minimum in the time axis and the frame corresponding to the trigger signal (around -175 ns) is the last one.
\begin{table}[h]
\centerline{\begin{tabular}{|l|l|c|}
\hline
July/August 2008  & Mini SDHCAL,\\
&3-12 GeV Pions PS beam CERN  \\
\hline
November 2008  & Mini SDHCAL,\\
&6 GeV Pions PS beam CERN  \\
\hline
June/July 2009  & Mini SDHCAL + 1 M$^{2}$,\\
&3-12 GeV Pions PS beam CERN \\
\hline
August 2009  & Mini SDHCAL +  1 M$^{2}$ + Absorber,\\
&10-150 GeV Pions SPS beam CERN \\
\hline
\end{tabular}}
\caption{Beam tests summary}
\label{Tab.beamtests}
\end{table}

\section{The mini SDHCAL prototype}
\subsection{Track reconstruction}
\label{Sec.track}
In each event we selected hits belonging to the same track using a time and position cut of respectively 400 ns (two internal clock periods) and 2 cm in radius.
Indeed, two hits belonging to two different layers (except the studied one) were used to build a track candidate. The track is confirmed if at least one hit is found in the third layer, than it is extended to the fourth layer and hits are searched with the same spatial and time criteria.

\subsection{Efficiency and multiplicity dependances on high voltage threshold}
\begin{figure}[h]
 \begin{center}
 \begin{minipage}{0.475\textwidth}
   \begin{centering}
     \includegraphics[width=1\textwidth,,height=5cm]{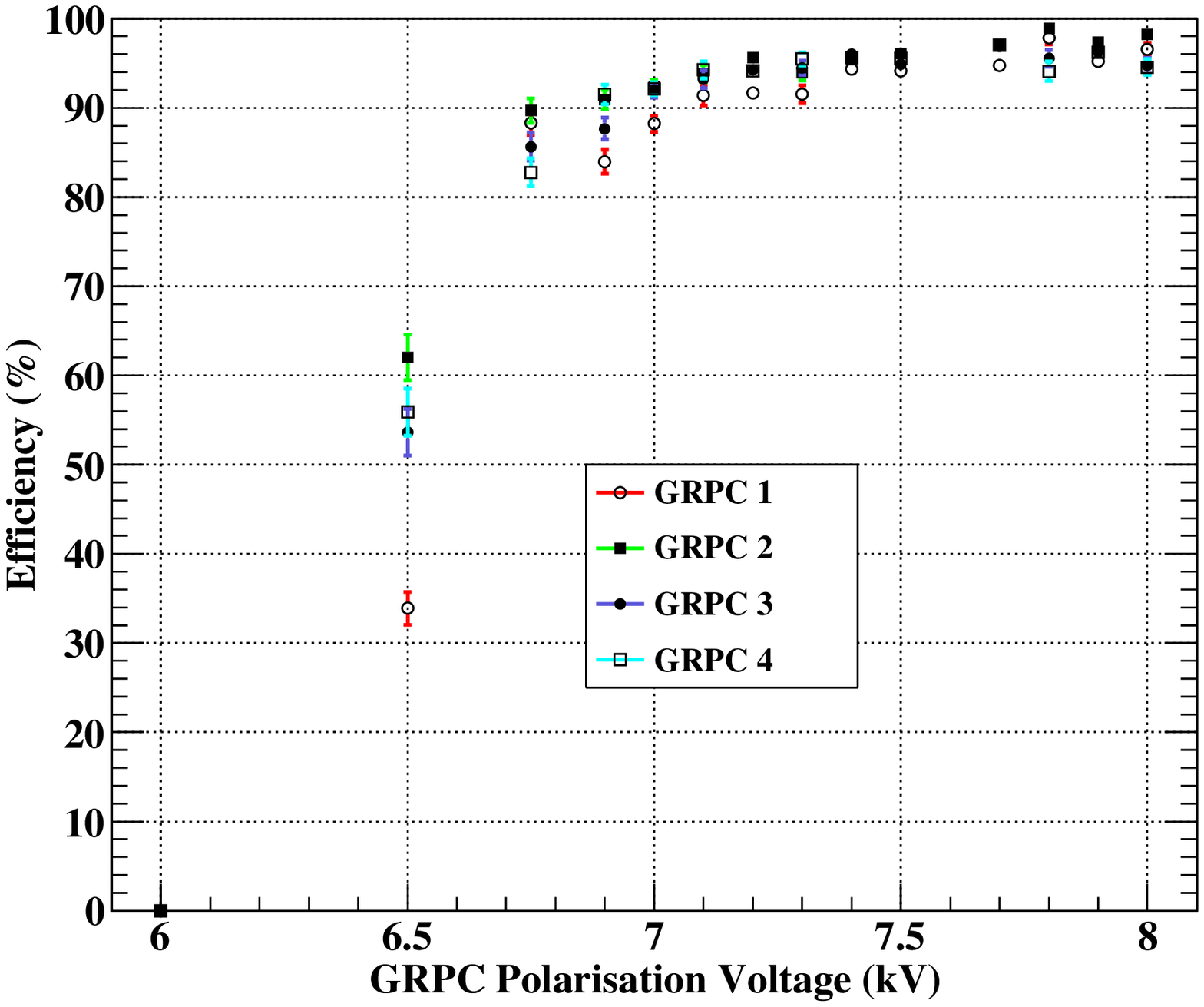}
   \end{centering}
 \end{minipage}
 \begin{minipage}{0.05\textwidth}
 \end{minipage}
 \begin{minipage}{0.475\textwidth}
   \begin{centering}
    \includegraphics[width=1\textwidth,height=5cm]{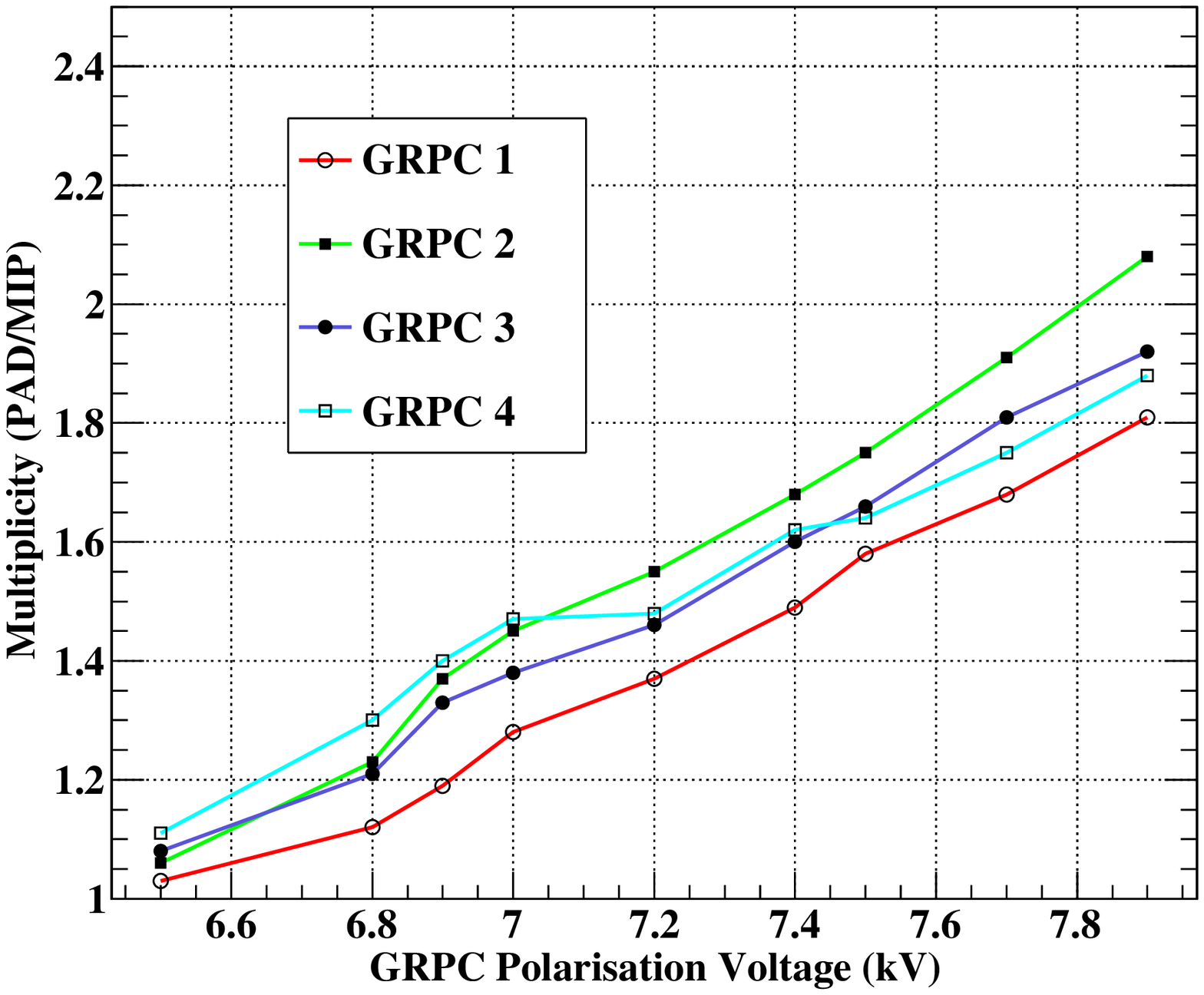}
   \end{centering}
 \end{minipage}
\caption{Efficiency vs high voltage (left) and multiplicity vs high voltage (right) for 4 small GRPCs.}
\label{Fig.hv} 
 \end{center}
\end{figure}

The choice of the electric field applied between the two glass plates is very important since it is responsible for the charge multiplication process. A high voltage scan was performed to study the detector performance.
The efficiency ($\epsilon$) of a layer is defined by the ratio between the number of tracks reconstructed in four layers and tracks reconstructed in the other three layers. The multiplicity ($\mu$) is the number of fired pads (leaving out inefficient tracks $\mu\ge$1). These two parameters were studied to define the optimal point with a low multiplicity and a high efficiency. The dependance of the efficiency and multiplicity on high voltage can bee seen in Figure \ref{Fig.hv} for a threshold of 165 fC. 
 A value of 7.4 kV high voltage polarisation is taken as standard working point with $\epsilon=95\%$ and $\mu=1.6$.

As for the high voltage study, a threshold scan was done with 7.4 kV high voltage to find the best triggering threshold wich gives a high efficiency and low multiplicity with minimal noise contribution. This study confirmed the choice of 165 fC as working point (no loss of efficiency and low noise).
\subsection{Detector homogeneity and stability in time}
\begin{figure}[!h]
   \begin{center}
    \includegraphics[width=1\textwidth,height=3cm]{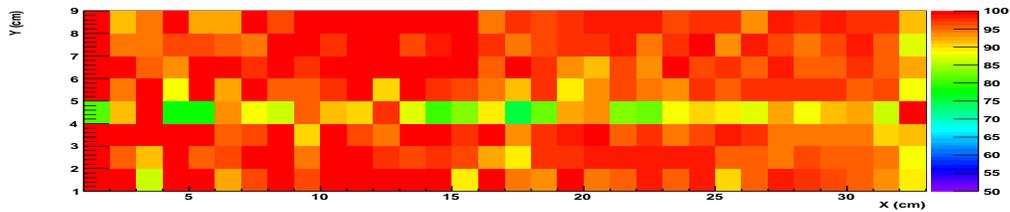}
   \caption{Efficiency map for one GRPC.}
\label{Fig.map} 
 \end{center}
\end{figure}

\begin{figure}[!h]
   \begin{center}
    \includegraphics[width=0.7\textwidth,height=5cm]{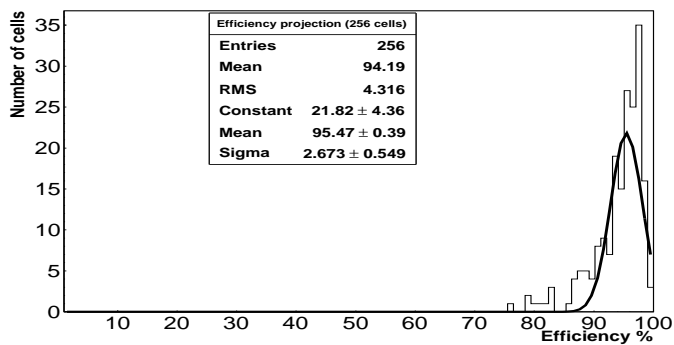}
   \caption{Projection of the efficiency map of one GRPC (256 cells).}
\label{Fig.disp} 
 \end{center}
\end{figure}

To study the homogeneity of the detector, an $(x-y)$ efficiency map was made using the tracking method explained in section~\ref{Sec.track}. Figure \ref{Fig.map} shows the efficiency map of one GRPC. The bulk uniformity dispersion including statistical error is less than $3\%$ (Figure \ref{Fig.disp}). Some systematical effects due to the edge and the fishing line separating the glass plates in the middle of the chamber appear in this efficiency map.

The stability in time of the GRPCs was checked over many runs with a fixed high voltage (7.4 kV) and threshold (165 fC). Cells were divided to three populations of efficient($\epsilon> 65\%$), medium ($20\% <\epsilon<65\%$) and dead cells($\epsilon< 20\%$).

The migration between populations was found to be less than $0.4\%$ with a fraction of dead cells of $0.1\%$. This demonstrates the good stability of the detector over a week.
 
\section{The large GRPC}

With the same electronics of the small GRPCs, a 1 m$^{2}$ GRPC was built at Lyon. It features 144 ASICs and 9216 $1\times1$ cm$^{2}$ cells. The fishing line separating the glass plates was replaced by small ceramic balls to reduce the contact surface and improve the gas circulation in the chamber.
A first attempt in beam tests failed due to some technical problems caused by connectors to the anode rendering the collected data was unusable. The results shown here were obtained with cosmic rays in the laboratory.

\begin{figure}[!h]
\begin {center}
\includegraphics[width=0.7\textwidth,height=5cm]{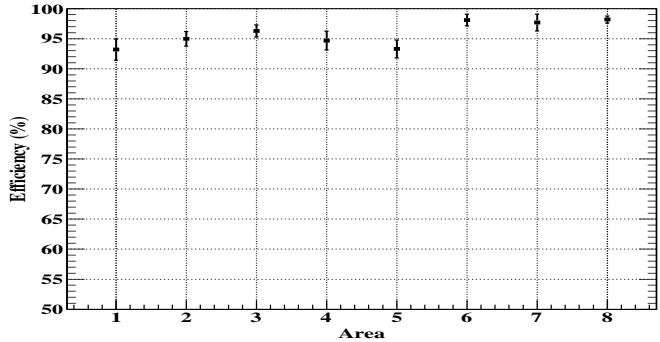}
\caption{Efficiency of different areas in the large GRPC.} 
\label{Fig.effsq}
\end {center}
\end{figure}

With a cut on time << to the trigger >> which represents the cosmics signal ($0<$Cosmic event time<1.2 $\mu $s), a noise contamination of 1\% was measured.
A position scan was performed to check the homogeneity of the large chamber. We divided it into 8 areas and calculated the average efficiency in each area. The result of this scan, shown in Figure~\ref{Fig.effsq}, proves a good homogeneity of the efficiency in the large GRPC. The multiplicity was also calculated and we found an average of 1.65 which is close to small GRPCs results.

\section{Conclusion}
A high efficiency coupled with low multiplicity is a key for a granular hadronic calorimeter. A set of small and large GRPCs were built and tested in pions beams and data analysis gives satisfying results, with a typical efficiency of 95\% and a multiplicity of 1.6, sufficient for an ILD hadronic calorimeter. 
\newpage
\section{Bibliography}

\bibliography{PFA}

\begin{thebibliography}{1}

\bibitem{Djouadi:2007ik}
G.~Aarons {\em et~al.}, ``{International Linear Collider Reference Design
  Report Volume 2: PHYSICS AT THE ILC},'' 2007.
\newblock \texttt {http://ilcdoc.linearcollider.org}.

\bibitem{Brient:2002gh}
J.-C. Brient and H.~Videau, ``{The calorimetry at the future e+ e- linear
  collider},'' in {\em Proceedings of APS / DPF / DPB Summer Study on the
  Future of Particle Physics (Snowmass 2001), Snowmass, Colorado, 30 Jun - 21
  Jul 2001, pp E3047.}, 2002.

\bibitem{Thomson:2009rp}
M.~A. Thomson, ``{Particle Flow Calorimetry and the PandoraPFA Algorithm},''
  {\em Nucl. Instrum. Meth.}, vol.~A611, pp.~25--40, 2009.

\bibitem{Group:2010eu}
T.~I.~C. Group, ``{The International Large Detector: Letter of Intent},'' 2010.
\newblock \texttt
  {http://www.ilcild.org/documents/ild-letter-of-intent/LOI\_final.pdf}.

\bibitem{Behnke:2007gj}
T.~Behnke, (Ed.~) {\em et~al.}, ``{ILC Reference Design Report Volume 4 -
  Detectors},'' 2007.
\newblock \texttt {http://ilcdoc.linearcollider.org}.

\bibitem{Bouchel:2007zz}
M.~Bouchel {\em et~al.}, ``{HARDROC1, readout chip of the digital hadronic
  calorimeter of ILC},''
\newblock CERN-2007-007, \texttt
  {http://cdsweb.cern.ch/record/1091461/files/p309.pdf}.

\end{thebibliography}




\bibliographystyle{ieeetr}
\end{document}